# Testing health expenditure rationing under universal healthcare coverage: the case of Italy

*Leonardo Becchetti, University of Tor Vergata, Rome*

*Nazaria Solferino, Universitas Mercatorum*

**Abstract**

We investigate the phenomenon of (private, out-of-pocket) "health expenditure rationing", or whether out-of-pocket health expenditures are shaped by income independently of actual health needs. Using microdata from an Italian Health Interview Survey, we assess the extent to which income explains variation in private health spending, after controlling for objective health conditions, self-assessed health, and a comprehensive set of socio-demographic factors. We find that individuals in the highest income brackets spend approximately €300 more annually than those in the lowest. Our findings point to a structural inequity in access and highlight the need for policy measures that address not only formal coverage but also the underlying role of income in shaping healthcare use.

1. **Background**

The relationship between income and healthcare access has become increasingly important as public healthcare systems across the world face growing challenges in meeting patient needs. In many countries, these challenges have led to longer waiting queues for treatment and, in many cases, limited access to necessary care. As a result, individuals are turning more frequently to out-of-pocket (OOP) healthcare expenses—direct payments made by individuals to access private health services—to avoid delays and receive care promptly. This shift from public to private, out-of-pocket spending raises serious concerns about health equity, especially for low-income households.

According to OECD Health Statistics, out-of-pocket (OOP) spending in Italy accounted for approximately 23% of total healthcare spending in the years 2021-2022 (De Bevus et al.,2024). This is higher than the EU average, which is closer to 1%. Similarly, in the United States, OOP costs represented 11.1% of total healthcare spending in 2022 (OECD, 2023).

If income is a key determinant of healthcare access, low-income households are particularly vulnerable, often delaying or forgoing necessary treatments due to financial limitations. A World Bank study found that in 2020, 8.6% of the global population faced catastrophic health expenditures, with this figure being particularly high in low- and middle-income countries with less robust healthcare systems (World Bank, 2021). In Europe, approximately 9.6% of Italians reported unmet healthcare needs due to cost in 2020, higher than the EU average of 7.2% (OECD, 2021). In the United States, nearly 24% of adults reported avoiding medical care in 2022 because of cost concerns (KFF, 2022).

The relationship between income and healthcare spending has been extensively studied in economic literature, emphasizing how lower-income groups face barriers to accessing necessary healthcare services. Theoretical models, such as Grossman's (1972) demand for health, highlight that health demand is investment in human capital, where individuals with higher incomes are better equipped to allocate resources toward healthcare needs. Newhouse (1977) builds on this by demonstrating that wealthier individuals typically have better access to high-quality healthcare services, primarily due to their greater ability to cover out-of-pocket (OOP) costs.

These findings emphasize that as income rises, so does the ability to consume healthcare services, suggesting a positive relationship between financial resources and access to care. However, for low-income households, these dynamics differ significantly. Research by Xu et al. (2003) and Wagstaff and van Doorslaer (2003) reveals that in contexts of financial constraints, healthcare spending is often deprioritized in favor of more immediate needs like food and housing. This leads to deferred treatments and increased exposure to catastrophic health expenditures, which in some cases push families into poverty or substantial debt. This is particularly evident in countries without comprehensive public healthcare systems, where individuals are forced to rely on private healthcare, leading to higher OOP costs. Along the same line Gertler and Gruber (2002) suggest that poor households are often unable to invest in preventive care due to immediate financial pressures, which leads to greater reliance on emergency care later on, thereby increasing overall health costs. This view is supported by Krishna (2004), who discusses how adverse health events can deepen existing poverty cycles, further entrenching inequality in health outcomes. Brekke and Kverndokk (2012) extend this argument by exploring the effects of income redistribution on health and socioeconomic inequality, noting that policies designed to equalize income without addressing health disparities may inadvertently widen the social gradient in health outcomes. Their findings emphasize that a holistic approach, which includes both income and health transfers, is necessary to reduce overall inequality and improve healthcare access across socioeconomic groups. Deaton and Paxson (1998) note that variability in income can force families to forgo necessary care, exacerbating existing health disparities and creating further barriers to equitable healthcare access. Skoufias and Quisumbing (2005) found that such shocks disproportionately affect low-income households, forcing them to cut back on healthcare expenditures, which in turn leads to worse health outcomes.

Cutler, Deaton, and Lleras-Muney (2006) highlight how socio-economic factors influence health behaviors and decision-making, arguing that financial insecurity leads to choices that negatively impact long-term health outcomes. According ti Banerjee and Duflo (2007) low-income households often postpone or avoid essential healthcare due to immediate financial constraints, despite the long-term costs of doing so. More recently, Berezowski et al. (2023) explored how healthcare rationing is becoming more prevalent as healthcare costs continue to rise. The authors argue that healthcare rationing, particularly in the context of resource scarcity, can sometimes result in essential treatments being withheld, thus exacerbating health disparities. The increasing reliance on rationing as a means to manage limited healthcare resources underscores the critical role of income in determining healthcare access. Meulman et al. (2025) enrich this perspective by quantifying the role of chronic conditions and social determinants in healthcare inequalities. Their analysis of Dutch data reveals that chronic illness and factors such as income security, education, and social capital account for nearly all disparities in self-rated health and a large portion of differences in healthcare expenditures, indicating that reducing health inequalities requires integrated action beyond the health sector.

Empirical evidence from health-income elasticity in high-income countries provides insight into the relationship between income and health spending. Acemoglu, Finkelstein, and Notowidigdo (2013), using exogenous variation in oil prices, found that income elasticity of health spending in the U.S. is positive but less than one (around 0.7), implying that rising income contributes only modestly to increased health spending. Similarly, Moscone and Tosetti (2010) observe that healthcare is a necessity rather than a luxury, suggesting a long-run elasticity well below one. These findings highlight that while income matters, other factors—such as public provision and structural determinants—play a critical role in shaping healthcare consumption. Piscopo, Groot, and Pavlova (2024), focusing on EU countries, show that GDP is a significant determinant of public health expenditure, while other factors—such as debt levels or political ideologies—have more modest or insignificant effects. This suggests that economic capacity, more than short-term political shifts, drives public health investment. Recent research (Cylus et al., 2024; Waitzberg et al., 2024) highlights that more equitable designs of co-payments and exemptions can significantly reduce the risk of catastrophic expenditures and mitigate regressivity in healthcare financing.

Insurance and health financing mechanisms play a critical role in mitigating the impact of income constraints. Jütting (2003) examines micro-insurance schemes as an effective tool to improve healthcare access for low-income populations by reducing the financial burden of OOP expenses. Similarly, O'Donnell et al. (2008) demonstrate that public insurance schemes can significantly enhance healthcare utilization by covering a substantial portion of medical costs, thus enabling lower-income individuals to access necessary care without financial strain (O'Donnell et al., 2008). Brenna (2025), focusing on Italy's National Health

System, finds that individuals with voluntary health insurance are more likely to use private healthcare—especially for specialist visits and diagnostics—suggesting that access to supplemental insurance can reproduce inequalities even within universal systems.

The literature clearly establishes that income constraints are a major barrier to equitable healthcare access. The rising costs of care and increased demands on public healthcare systems have made (public complimentary) healthcare rationing a necessary practice, but one that raises significant concerns about equity of access to healthcare, particularly for lower-income households (Berezowski et al., 2023; Scheunemann & White, 2011).

Our paper makes a significant and original contribution to this literature by investigating the phenomenon of health expenditure rationing through the interplay of income, health status, and out-of-pocket spending within a universal healthcare system. Unlike prior studies, which often focus either on income disparities or healthcare access separately, our research uniquely combines detailed individual-level data on income, self-assessed health (SAH), and diagnosed pathologies to test whether out-of-pocket expenditures are disproportionately influenced by income, independent of actual health needs. This allows us to highlight the paradoxical finding that higher-income individuals spend more out-of-pocket despite reporting fewer health problems, revealing an implicit rationing mechanism based primarily on economic capacity rather than medical necessity.

The use of SAH in our research is important as it captures subjective and latent aspects of health conditions that standard medical diagnoses might overlook or delay, such as early symptoms, psychological wellbeing, and behavioral factors influencing health outcomes. This forward-looking measure strengthens our analysis by providing a more comprehensive and nuanced understanding of an individual's true health status, which is crucial when assessing whether healthcare spending aligns with medical necessity or economic constraints.

Furthermore, our empirical approach incorporates extensive controls for key socio-demographic variables—such as education, employment status, gender, and family composition—that are known to influence both health outcomes and healthcare utilization. Controlling for these factors ensures that the positive effect of income on out-of-pocket expenditure is not spuriously driven by correlated social characteristics, thus increasing the robustness and credibility of our findings.

Additionally, by focusing on Italy, a country with universal healthcare coverage but heterogeneous access to private insurance and varying degrees of private-public care interaction, our study fills an important empirical gap in the literature. We provide novel evidence on how income inequality translates into financial barriers and implicit rationing, even in systems designed to guarantee equitable access. Finally, our analysis explicitly quantifies the downward bias in estimating health expenditure rationing that arises when failing to consider differential private insurance coverage and health heterogeneity across income groups. This methodological refinement offers valuable insights for future research and policymaking aimed at accurately identifying and addressing inequities in healthcare financing.

## 2. Methods

### 2.1. Research Questions

In this study we explore how income shapes healthcare spending within a formally universal healthcare system, focusing on whether private health expenditures reflect genuine medical need or simply financial ability. Leveraging both objective (diagnosed conditions) and subjective (self-assessed) health indicators, we aim to disentangle the extent to which income drives spending independently of health status. We therefore wonder whether those with higher incomes spend more simply because they can, while individuals in worse health may spend less due to financial barriers. This issue is central to understanding equity in healthcare systems that formally guarantee universal access. Prior studies (e.g., Gertler & Gruber, 2002; Acemoglu et al., 2013; Meulman et al., 2025) suggest that health-related consumption may not align with actual need when income constraints are binding

We therefore formulate the following research hypothesis

*H01: does higher income lead to greater out-of-pocket healthcare expenditure independently of health needs, and to what extent does this reflect implicit rationing among lower-income individuals?*

## 2.2. Model Specification

The benchmark specification estimated in the paper is

$$OPath_{d,i} + \alpha_1 BMI_i + \alpha_2 Smoker_i + \alpha_3 Alcohol_i + \alpha_4 SportAct_i + \alpha_5 Female_i + \alpha_6 Age_i + \alpha_7 Age_i^2 + \sum_d \theta_d DEducation_{d,i} + \alpha_4 Children$$

where the dependent variable (OOP_HEXP)_ is out of pocket health expenditure of the i-th individual, calculated as yearly health expenditure considering unreimbursed costs for prevention and treatment, such as medicines, paramedical products and medical visits, transportation costs, etc.

Our main regressors of interest are income level dummies and our null hypothesis is not rejected if $\sum_a \beta_a = 0$. Health control variables include dummies mesuring levels of self-assessed-health, a list of pathologies (heart problems, hypertension, lung diseases, cancer, artritis, asthma, diabetes), body mass index and (0/1) dummies capturing whether the respondent smokes, drinks alcohol or practices sports. Among socio-demographic controls we include a gender dummy, age and age squared, education level dummies, a dummy for marital status for those living with partner, the number of children, work status dummies, and 20-1 (0/1) regional dummies picking up regional effects, particularly relevant in the Italian case given that health is managed at regional level.

The model is estimated on data from a survey on a sample of Italians run by Next. The survey is representative of the Italian population per gender, education and geographical area.

3.Results

Our main variable of interest, out-of-pocket health expenditure, is calculated as yearly health expenditure considering unreimbursed costs for prevention and treatment, such as medicines, paramedical products and medical visits, transportation costs, etc. As a consequence, health costs anticipated and reimbursed by insurance are not included here.

About income distribution around 21 percent of sample respondents declare income below 12 thousand euros. Only 10 percent of respondents are above 30 thousand euros and in the three highest income brackets. In our sample, low-income individuals appear slightly underrepresented as 21% report earning less than €12,000 annually, compared to 36% of taxpayers according to MEF (2022). Conversely, MEF reported that 11,5% are in the €35,000–€75,000 income range, suggesting good alignment in the upper brackets.

Average health expenditure is 690 for individuals with income between 50.601 and 70.400 euros, while 391 for those in the lowest declared income bracket (below 11,700 euros), showing that the out-of-pocket health expenditure gap between the two classes is around 300 euros. the difference is significant using 95 percent confidence intervals (Table 1, panel 1). The gap is likely to be higher if we consider that higher income individuals are more likely to have a private insurance that covers part of their health expenditure.

One might wonder whether respondents with higher income spend more out of pocket because their health is worse. We however find that it not the case, while exactly the opposite. Pathologies reported in our survey are: heart problems, hypertension, lung diseases, cancer, artitis, asthma, diabetes. The average number of pathologies for individuals in the next to highest (50.601-70.400) income bracket is 0.6, while that for individuals in the lowest income bracket is 1.6 (Table 1, panel 2). When looking at single pathologies we find that 27 percent (44 percent) of respondents in the lowest income bracket declare they have diabetes (hypertension) while none of those in the highest income bracket (and in any case 9.8 percent among those in the last three highest income brackets (income higher than 30,000 euros) suffer from diabetes and 24 percent from hypertension) (Table 1, panels 3-4).

We check whether these findings are confirmed when controlling for self-assessed health (SAH). As we know from the literature this variable provides a forward-looking measure of individual health status, capturing early signals and subjective experiences that formal diagnoses may miss or delay. SAH significantly predicts future insurgence of chronic diseases, even after accounting for sociodemographic, lifestyle, and health system variables (Becchetti et al., 2018). This suggests that SAH reflects both observable and latent health conditions. Prior research has also shown SAH to be a strong predictor of mortality and morbidity across diverse populations (Idler & Kasl, 1995; Doiron et al., 2015). SAH can account for psychological and behavioral factors that affect health but are not captured by objective medical records. Therefore, relying solely on diagnosed pathologies underestimates the complexity and evolution of health conditions, while SAH offers crucial early-warning insight for health policy and preventive care. Differences in self-assessed health between low and high income individuals in our sample are huge. Lowest income individuals give responses mainly between one and three (very bad, discrete, good) while those in the highest income bracket between three and five (good, very good, excellent) (Figure 2).

To sum up the observed out of pocket expenditure gap is around 300 euros per year but it highly likely to be a downward biased estimate of health expenditure rationing if we consider heterogeneity in access to private insurance reimbursement which grows with income and the health differences who are in favour of the richer.

To test our research hypothesis and see whether the observed descriptive differences hold when controlling for all observable concurring factors we estimate the model described in section 3

Empirical findings show that income has a strongly positive and significant effect on the dependent variable thereby rejecting the null hypothesis of no health rationing (Table 3). The magnitude of the effect is such that being in the highest (50.601-70.400) income bracket reduces by 22 percent the probability of declaring the lowest out of pocket health expenditure versus being in the lowest income bracket.

As identification strategy to test whether the observed significant correlation between income and out of pocket health expenditure can be considered a direct causal nexus we choose the regional/gender specific average income as instrument.(see, e.g., Angrist & Imbens, 1995; Currie & Moretti, 2003; Chetty et al., 2014, for similar uses of group-level income instruments). We expect the instrument to be relevant since individual respondent's income should be correlated with this average. We expect it to be valid since, for large enough groups as in our case, the average variable should capture ex ante (cohort effect or human capital) variation not affecting per se our dependent variable. It is in fact reasonable to believe that average income/gender group income cannot affect the individual out of pocket expenditure.

For a parsimonious IV estimate we slightly modify our model using OLS approach and considering income classes as a continuous income variables which we instrument with the average regional/gender sample income.

First stage estimates confirm that the instrument is relevant and not weak (First-Stage F-statistic 10.48). These tests are robust to weak instruments. Second stage estimates show that the instrument significantly affects our dependent variable.

We as well provide a falsification test where the instrument replaces the instrumented variable in an OLS estimate of our benchmark specification, estimated in the subsample of the lowest admissible level of the instrument (lowest income group). The instrument is not significant thereby not rejecting the hypothesis that the instrument affects the dependent variable only through the instrumented variable. We repeat our estimate in sample splits according to gender, education and geographical area. The nexus between income and out of pocket health expenditure remains significant and growing in income classes with higher values for those living in the South of Italy and females. The elders have an insignificant effect in the highest income decile due to the very low numerosity of this class for them (Table 2, columns 1-6). The problem disappears when we cumulate the highest two income classes

# 4. Discussions

Results presented above highlight the significant role that income constraints play in shaping healthcare expenditures, particularly for vulnerable populations.

A central finding of our analysis is the paradox that higher-income individuals spend more on health care out-of-pocket, yet report fewer pathologies. Meanwhile, those with lower income, despite having greater health needs, spend significantly less. This pattern reflects a form of implicit rationing based on economic capacity, rather than medical necessity. This distortion is likely to generate three inequitable dynamics. First, low-income individuals, who are on average more burdened by illness, often delay or forgo necessary care due to financial constraints. This leads to worse health outcomes and higher long-term costs for the public health system. Second, wealthier individuals, though they primarily rely on private providers, often rely on the public system as a complement or fallback, even when not strictly necessary. As an example: i) someone may consult a specialist privately, but then schedule public diagnostic tests (e.g., MRI) to avoid high out-of-pocket fees. Parents may use private pediatric care but access public facilities for school certifications or vaccine records. Patients may book a public consultation (e.g., in intramoenia) after already receiving a private opinion, simply to access prescriptions or confirm diagnoses. In these cases, the public system is used not out of clinical urgency, but to optimize costs or convenience, contributing to overcrowding and delaying care for those with no alternatives.

Finally, high-income users willing to pay premium fees also distort the pricing equilibrium of the private market as providers are incentivized to maintain high tariffs, prices remain structurally unaffordable for low- and middle-income groups, and even semi-public channels like intramoenia tend to reflect market-driven rates, rather than equitable access. The result is a dual system: formal universality coexists with de facto exclusion for low income household when cumulating waiting lists in the public sector and financial barriers in the private sector, with those most in need accessing care less frequently, less promptly, and often too late.

Notice that, we do not suggest that there is necessarily an inherent injustice, but the evidence indicates an implicit economic rationing of healthcare access. Individuals with fewer financial resources tend to use healthcare services less, even when facing clear medical needs. Our analysis focuses on serious conditions—such as diabetes, cardiovascular diseases, and arthritis, to capture situations of genuine clinical necessity rather than general consumption.

The data show that lower-income individuals experience more severe illnesses yet spend less on healthcare than wealthier individuals with comparable needs. This cannot simply be attributed to discretionary or non-essential consumption by the rich; rather, it reflects that those with greater need but fewer resources still have reduced access to care. Thus, even within a universal healthcare system, effective access remains influenced by income. Moreover, higher private spending by the wealthy can indirectly increase prices and divert resources from the public sector, further constraining access for those with limited means. While this may not demonstrate direct inequity, it underscores that the system implicitly rations care in an economically selective manner.

From a policy perspective, our study suggests several key interventions to reduce the financial burden of healthcare, particularly for low-income groups. First, expanding public health insurance coverage and introducing subsidies targeted at low-income households could help alleviate out-of-pocket costs and reduce financial barriers to care. For example, increasing the coverage of essential health services or reducing co-pays could significantly lower the out-of-pocket expenditures for lower-income families. Furthermore, addressing regional disparities is crucial. Policymakers should focus on improving healthcare infrastructure, including digital access and transportation, to ensure that people in underserved regions can access affordable care. Regional differences, such as those between the North and South of Italy, underscore the need for tailored policies that address local healthcare access issues.

Additionally, improving health literacy and providing financial support for preventive healthcare could help mitigate the negative effects of economic shocks. Finally, more effective healthcare rationing policies should be designed to prioritize equity and prevent worsening health outcomes for the most disadvantaged groups. By adopting these policies, Italy can work towards a healthcare system that ensures more equitable access for all, regardless of income or socioeconomic status.

## 5. Conclusions

We provide robust evidence of income-related inequalities in healthcare access within a formally universal healthcare system such as Italy. Our empirical analysis shows that out-of-pocket (OOP) healthcare expenditures are significantly and positively associated with income. On average, individuals in the highest income brackets spend approximately €300 more per year on healthcare than those in the lowest, despite being in significantly better health. This gap persists even after accounting for diagnosed pathologies, self-assessed health (SAH), and insurance coverage, strongly suggesting the presence of implicit health expenditure rationing based on economic capacity.

The analysis reveals that the higher-income groups are, on average, healthier across multiple indicators. They report fewer chronic conditions (such as diabetes and hypertension), and have significantly better self-assessed health. At the same time, these groups are more likely to hold private insurance policies that partially reimburse their health spending, which is not captured in our OOP measure, further reinforcing the concern that the €300 gap underestimates the true inequality.

We find that the probability of declaring the lowest OOP expenditure is 22% lower for individuals in the highest income group compared to those in the lowest, controlling for all observable health and demographic characteristics. These patterns suggest that those with higher income are better positioned to access healthcare services, even when need is lower, while those in worse health, but with fewer financial resources, can be forced to postpone or forego treatment due to affordability issues.

Our findings contribute to the growing literature on healthcare inequity under universal systems (e.g., Wagstaff & van Doorslaer, 2003; Meulman et al., 2025), offering novel micro-level evidence for Italy and highlighting a form of "silent" rationing, where formal access rights are not matched by effective utilization due to time barriers in the public system and financial barriers in the private system. They also reinforce concerns that income, education, employment status, and family composition mediate access to care in ways that policy often fails to acknowledge.

These insights carry strong policy implications. First, formal universality is not sufficient to guarantee equitable access to healthcare. Policymakers should recognize and address the hidden costs that discourage care-seeking behavior among low-income individuals, including waiting lists depending on the excess demand of a good sold below market price, co-payments, medication costs, and non-medical expenses like transportation. Second, insurance design matters: increasing the availability of comprehensive public or subsidized private insurance—especially for vulnerable populations—could reduce health rationing based on economic capacity. Third, preventive care programs should be more actively targeted at lower-income groups, who are both more at risk and more likely to delay care facing financial barriers to them. Finally, policy tools that integrate healthcare support with broader social protection mechanisms (such as income supplementation or housing assistance) may be more effective in reducing health-related inequalities than health-sector interventions alone.

Our results make clear that economic capacity remains a fundamental determinant of healthcare access, even in systems that aspire to equity. This form of implicit rationing undermines the principle of need-based care and calls for renewed attention to the financial realities faced by low-income households. Without targeted intervention, universal coverage risks remaining an abstract ideal rather than a lived reality.

## References

Acemoglu, D., Finkelstein, A., & Notowidigdo, M. J. (2013). Income and health spending: Evidence from oil price shocks. *The Review of Economics and Statistics*, 95(4), 1079–1095.

Angrist, J. D., & Imbens, G. W. (1995). Two-stage least squares estimation of average causal effects in models with variable treatment intensity. *Journal of the American Statistical Association,* 90(430), 431–442

**Figure 1 Out-of-pocket health expenditure, number of pathologies, hypertension and diabetes incidence for different income classes**

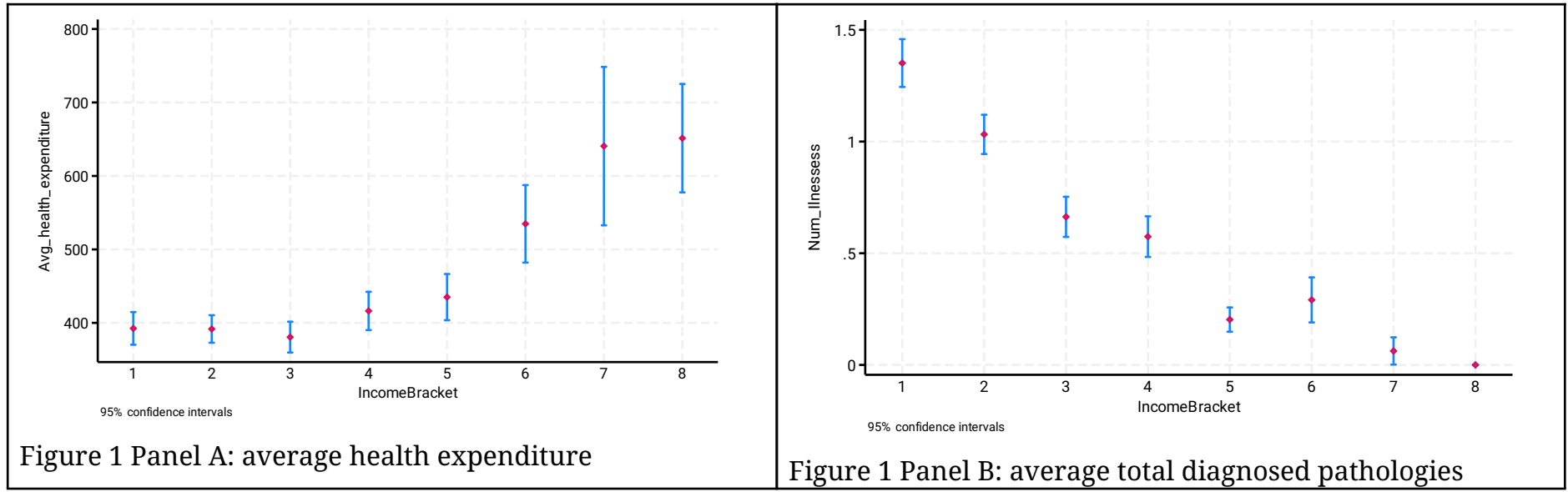

Figure 1 Panel A: average health expenditure

Figure 1 Panel B: average total diagnosed pathologies

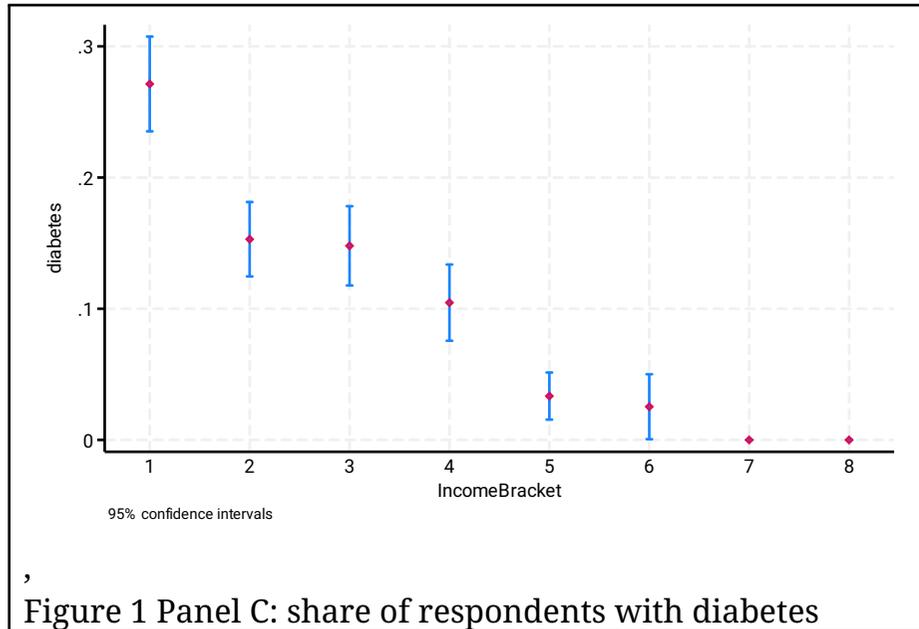

Figure 1 Panel C: share of respondents with diabetes

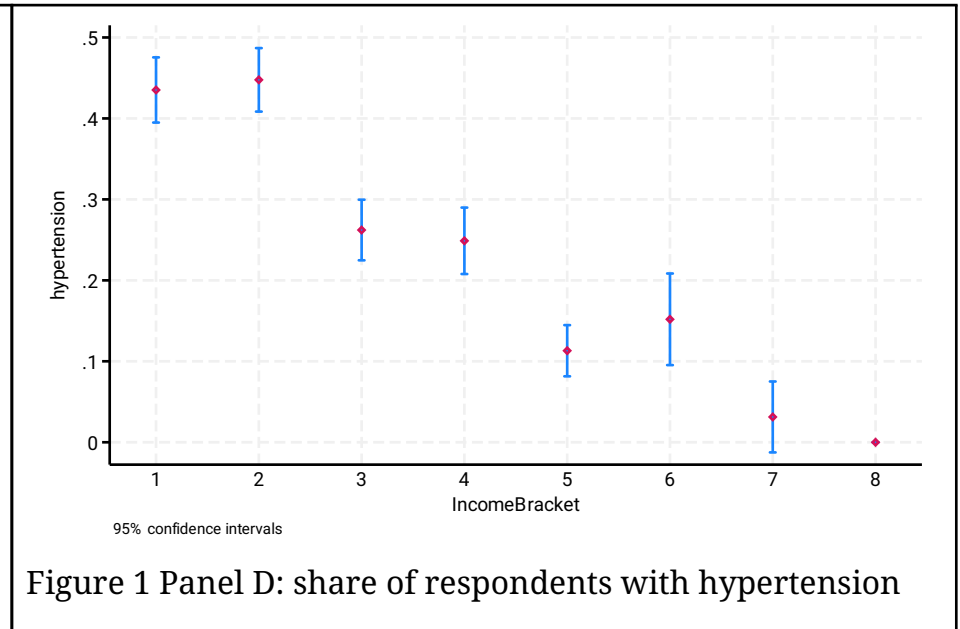

Figure 1 Panel D: share of respondents with hypertension

Income bracket 1= up to 12,000 euros , Income bracket 2= 12,001-15000 euros, Income bracket 3=15,000-20,000 euros, Income bracket 4=20,00-25,000 euros, Income bracket 5=25,000-30,000 euros , Income bracket 6=30,000-35,000 euros , Income bracket 7=35,000-42,000 euros , Income bracket 8= above 42,00 euros,

**Figure 2. Self assessed health for lowest income and income above bracket 6**

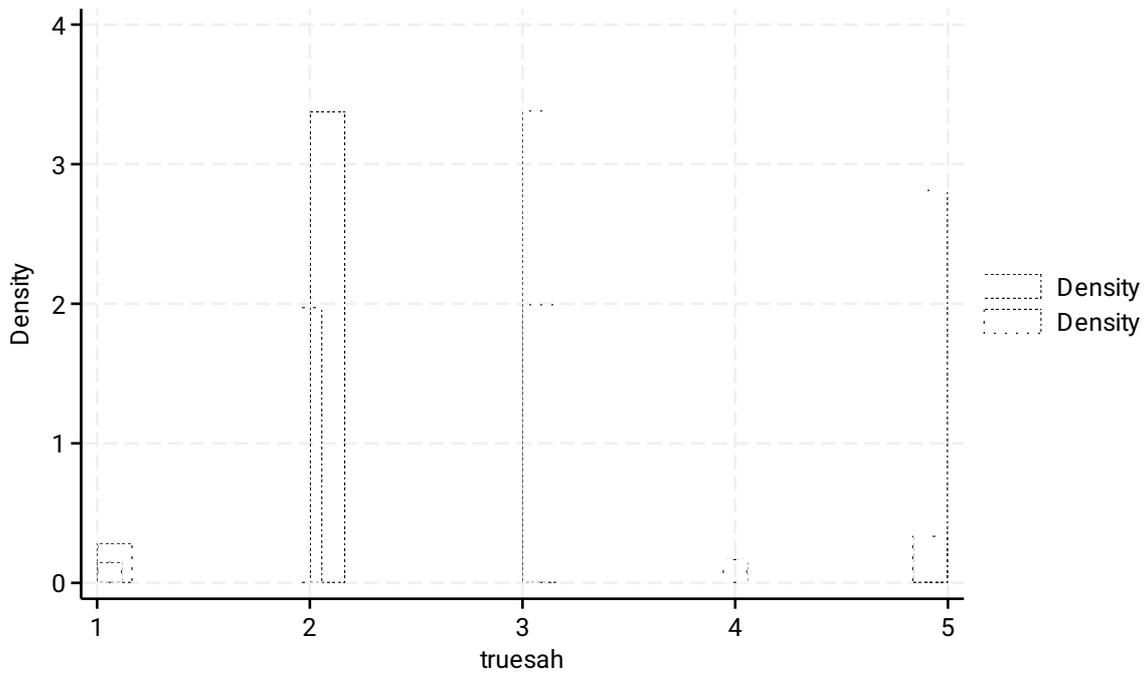

Legend: Self-Assessed-Health. 1: Very bad; 2: Discrete; 3: Good; 4: Very Good; 5: Optimal. Pink: lowest income. Green: income above sixth bracket

## Table 1 The effect of income on out of pocket health expenditure

| VARIABLES | (1) | (2) | (3) | (4) | IV First stage Dep var: continuous income | IV Second stage | Falsification test |
|---|---|---|---|---|---|---|---|
| Continuous income | | | | | | 2.93** (1.44) | 0.79 (0.001) |
| Average regional/gender income (1000 euros) | | | | | 0.083 (0.026) | | |
| Income 2 | -0.0304 (0.0574) | 0.0163 (0.0597) | 0.00320 (0.0610) | 0.0157 (0.0616) | | | |
| Income 3 | 0.0286 (0.0688) | 0.0552 (0.0721) | -0.00640 (0.0740) | -0.00498 (0.0742) | | | |
| Income 4 | 0.264*** (0.0718) | 0.247*** (0.0738) | 0.258*** (0.0758) | 0.251*** (0.0758) | | | |
| Income 5 | 0.441*** (0.0818) | 0.497*** (0.0834) | 0.516*** (0.0832) | 0.496*** (0.0824) | | | |
| Income 6 | 0.709*** (0.106) | 0.692*** (0.107) | 0.705*** (0.107) | 0.674*** (0.107) | | | |
| Income 7 | 0.964*** (0.178) | 1.041*** (0.184) | 0.980*** (0.185) | 0.918*** (0.183) | | | |
| Income 8 | 1.081*** (0.139) | 1.159*** (0.143) | 0.958*** (0.149) | 0.875*** (0.151) | | | |
| Female | -0.0314 | 0.00774 | -0.110** | -0.105** | | | |

|                       | (0.0431)     | (0.0447)     | (0.0488)     | (0.0493)     |
|-----------------------|--------------|--------------|--------------|--------------|
| Age                   | 0.0538***    | 0.0408***    | 0.0462***    | 0.0528***    |
|                       | (0.0110)     | (0.0112)     | (0.0113)     | (0.0116)     |
| $[Age]^2$             | -0.000300*** | -0.000245**  | -0.000302*** | -0.000354*** |
|                       | (0.000106)   | (0.000108)   | (0.000109)   | (0.000111)   |
| Secondary education   | 0.110*       | 0.313***     | 0.258***     | 0.276***     |
|                       | (0.0614)     | (0.0669)     | (0.0697)     | (0.0711)     |
| Tertiary education    | 0.423***     | 0.689***     | 0.614***     | 0.622***     |
|                       | (0.0888)     | (0.0956)     | (0.0983)     | (0.0995)     |
| Children              | 0.0578       | 0.159**      | 0.0758       | 0.0541       |
|                       | (0.0795)     | (0.0808)     | (0.0826)     | (0.0836)     |
| Stable relationship   | -0.114*      | -0.0204      | -0.0276      | -0.0367      |
|                       | (0.0621)     | (0.0643)     | (0.0663)     | (0.0670)     |
| Houseworker           | 0.0544       | 0.111        | 0.0639       | 0.00135      |
|                       | (0.287)      | (0.306)      | (0.313)      | (0.319)      |
| Retired               | 0.383**      | 0.101        | 0.0590       | 0.0364       |
|                       | (0.194)      | (0.207)      | (0.211)      | (0.212)      |
| Employed              | 0.359*       | 0.363*       | 0.332        | 0.338*       |
|                       | (0.186)      | (0.199)      | (0.205)      | (0.206)      |
| Unemployed            | 0.120        | 0.174        | 0.188        | 0.178        |
|                       | (0.177)      | (0.193)      | (0.199)      | (0.200)      |
| Ischemic heart disease |             | 0.401***     | 0.365***     | 0.373***     |
|                       |              | (0.102)      | (0.102)      | (0.104)      |
| Hypertension          |              | 0.512***     | 0.543***     | 0.530***     |
|                       |              | (0.0534)     | (0.0550)     | (0.0582)     |
| Lung Disease          |              | 0.784***     | 0.697***     | 0.694***     |
|                       |              | (0.122)      | (0.122)      | (0.126)      |
| Cancer                |              | 0.690***     | 0.578***     | 0.643***     |
|                       |              | (0.185)      | (0.184)      | (0.195)      |
| Artritis              |              | 0.329***     | 0.314***     | 0.298***     |
|                       |              | (0.0536)     | (0.0547)     | (0.0566)     |
| Asthma                |              | 0.425***     | 0.420***     | 0.427***     |
|                               | (0.0883) | (0.0890) | (0.0893) |          |
| fibro1cistica                 |          | -0.110   | -0.231   | -0.263   |
|                               |          | (0.219)  | (0.217)  | (0.218)  |
| Diabetes                      |          | 0.340*** | 0.348*** | 0.331*** |
|                               |          | (0.0577) | (0.0586) | (0.0593) |
| Bmi                           |          |          | -0.0159**| -0.0158**|
|                               |          |          | (0.00733)| (0.00763)|
| Smoker                        |          |          | -0.154***| -0.125** |
|                               |          |          | (0.0492) | (0.0497) |
| Drinking alcohol              |          |          | -0.134***| -0.128***|
|                               |          |          | (0.0192) | (0.0192) |
| Sport practice                |          |          | 0.0289   | 0.0182   |
|                               |          |          | (0.0227) | (0.0230) |
| Self-Assessed-Health (discrete)|         |          |          | 0.198    |
|                               |          |          |          | (0.173)  |
| Self-Assessed-Health (good)   |          |          |          | -0.0312  |
|                               |          |          |          | (0.184)  |
| Self-Assessed-Health (very good)|        |          |          | 0.422*   |
|                               |          |          |          | (0.249)  |
| Self-Assessed-Health (excellent)|        |          |          | 0.247    |
|                               |          |          |          | (0.194)  |
| Regional fixed effects        | Yes      | Yes      | Yes      | Yes      |
| /cut1                         | 3.053*** | 3.115*** | 2.331*** | 2.592*** |
|                               | (0.325)  | (0.339)  | (0.396)  | (0.441)  |
| /cut2                         | 4.353*** | 4.513*** | 3.760*** | 4.031*** |
|                               | (0.327)  | (0.342)  | (0.400)  | (0.445)  |
| /cut3                         | 5.440*** | 5.675*** | 4.952*** | 5.224*** |
|                               | (0.342)  | (0.358)  | (0.413)  | (0.467)  |
| /cut4                         | 5.974*** | 6.240*** | 5.517*** | 5.779*** |
|                               | (0.382)  | (0.406)  | (0.456)  | (0.496)  |
| Log likelihood                | -3217.61 | -3028.76 | -2982.18 | -2963.77 |

| | | | | | | |
|---|---|---|---|---|---|---|
| Adj. R Squared | | 0.10 | 0.15 | 0.17 | 0.17 | |
| F(1, 2805) = | | | | | 10.48 (0.001) | |
| F( 50, 2802) = | | | | | | 9.63 (0.00) |
| Observations | | 5,359 | 5,359 | 5,359 | 5,359 | |

Omitted benchmark: lowest income level, student, single, no children, primary education, self-assessed-health: bad. Robust standard errors in parentheses. *** p<0.01, ** p<0.05, * p<0.1.

**Table 2. The effect of income on out-of-pocket health expenditure - Sample splits**

| VARIABLES | (1) Above 54 | (2) Below 55 | (3) South and isles | (4) South and isles excluded | (5) Female | (6) Male |
|---|---|---|---|---|---|---|
| Income 2 | 0.0889 | -0.0795 | -0.203** | 0.188** | 0.0768 | -0.0622 |
| | (0.0719) | (0.143) | (0.0974) | (0.0825) | (0.0864) | (0.0888) |
| Income 3 | -0.189** | 0.273** | 0.0238 | -0.0283 | -0.159 | 0.109 |
| | (0.0940) | (0.120) | (0.113) | (0.0978) | (0.104) | (0.109) |
| Income 4 | 0.223** | 0.366*** | 0.322*** | 0.227** | 0.303*** | 0.199* |
| | (0.107) | (0.111) | (0.125) | (0.0985) | (0.107) | (0.112) |
| Income 5 | 0.505*** | 0.536*** | 0.113 | 0.671*** | 0.526*** | 0.484*** |
| | (0.155) | (0.102) | (0.156) | (0.100) | (0.128) | (0.112) |
| Income 6 | 0.654*** | 0.706*** | 0.619*** | 0.723*** | 0.762*** | 0.641*** |
| | (0.202) | (0.131) | (0.205) | (0.127) | (0.166) | (0.144) |
| Income 7 | 1.249*** | 0.864*** | 0.990*** | 0.931*** | 0.856*** | 0.842*** |
| | (0.389) | (0.211) | (0.353) | (0.215) | (0.313) | (0.223) |
| Income 8 | 0.400 | 1.036*** | 1.256*** | 0.802*** | 1.122*** | 0.624*** |
| | (0.377) | (0.173) | (0.272) | (0.178) | (0.424) | (0.174) |
| Age | 0.200** | 0.0584* | 0.0387** | 0.0612*** | 0.0435*** | 0.0618*** |
| | (0.0839) | (0.0330) | (0.0188) | (0.0148) | (0.0160) | (0.0175) |
| $[Age]^2$ | -0.00130** | -0.000416 | -0.000122 | -0.000477*** | -0.000287* | -0.000428** |
| | (0.000597) | (0.000416) | (0.000181) | (0.000141) | (0.000154) | (0.000166) |
| Children | -0.0374 | 0.129 | -0.0811 | 0.149 | -0.00695 | 0.116 |
| | (0.142) | (0.109) | (0.134) | (0.110) | (0.123) | (0.114) |
| Stable relationship | 0.0990 | -0.183* | 0.217** | -0.201** | -0.000402 | -0.112 |
| | (0.104) | (0.0950) | (0.104) | (0.0881) | (0.0927) | (0.104) |
| Houseworker | 3.950*** | 0.519* | -0.447 | 0.460 | -0.461 | 0.0200 |
| | (0.244) | (0.311) | (0.277) | (0.399) | (0.584) | (0.248) |
| Retired | 4.501*** | 0.332 | 0.0404 | 0.694* | -0.243 | |
| | (0.221) | (0.227) | (0.258) | (0.398) | (0.571) | |
| Employed | 4.442*** | 0.0687 | 0.150 | 0.379 | -0.537 | 0.243 |
| | (0.211) | (0.206) | (0.250) | (0.392) | (0.565) | (0.232) |
| Unemployed | 3.950*** | 0.519* | -0.447 | 0.460 | -0.461 | 0.0200 |
| | (0.244) | (0.311) | (0.277) | (0.399) | (0.584) | (0.248) |
| Female | -0.182*** | -0.0355 | 0.0463 | -0.209*** | | |
| | (0.0700) | (0.0740) | (0.0792) | (0.0639) | | |
| Secondary education | 0.149 | 0.478*** | 0.366*** | 0.232** | 0.435*** | 0.0314 |
| | (0.0937) | (0.143) | (0.107) | (0.0950) | (0.0963) | (0.106) |
| Tertiary education | 0.663*** | 0.809*** | 0.672*** | 0.644*** | 0.494*** | 0.565*** |
| | (0.192) | (0.163) | (0.164) | (0.128) | (0.150) | (0.139) |
| Ischemic heart disease | 0.385*** | 0.442 | 0.814*** | 0.127 | 0.411** | 0.329** |
| | (0.111) | (0.346) | (0.170) | (0.133) | (0.187) | (0.130) |
| Hypertension | 0.595*** | 0.408*** | 0.471*** | 0.598*** | 0.569*** | 0.512*** |
| | (0.0679) | (0.135) | (0.0880) | (0.0778) | (0.0789) | (0.0875) |
| Lung Disease | 0.769*** | -0.0308 | 0.705*** | 0.732*** | 0.681*** | 0.552*** |
| | (0.136) | (0.393) | (0.201) | (0.160) | (0.224) | (0.154) |
| Cancer | 0.846*** | -0.0513 | 0.327 | 0.876*** | 0.581** | 0.905** |
| | (0.228) | (0.336) | (0.268) | (0.253) | (0.242) | (0.364) |
| Artritis | 0.297*** | 0.560*** | 0.362*** | 0.270*** | 0.458*** | 0.0922 |
| | (0.0616) | (0.178) | (0.0891) | (0.0728) | (0.0732) | (0.0952) |
| Asthma | 0.315*** | 0.722*** | 0.420*** | 0.429*** | 0.232* | 0.582*** |
| | (0.102) | (0.174) | (0.144) | (0.113) | (0.127) | (0.124) |
| fibro1cistica | 0.711** | -1.114*** | -0.326 | -0.254 | -0.508** | 1.732*** |
| | (0.293) | (0.406) | (0.338) | (0.305) | (0.246) | (0.201) |
| Diabetes | 0.324*** | 0.567*** | 0.419*** | 0.286*** | 0.376*** | 0.289*** |
| | (0.0638) | (0.206) | (0.0953) | (0.0778) | (0.0789) | (0.0977) |
| Bmi | -0.0120 | -0.0307*** | -0.00550 | -0.0315*** | -0.00400 | -0.0559*** |
| | (0.00946) | (0.0116) | (0.00679) | (0.00970) | (0.00435) | (0.0129) |
| Smoker | -0.239*** | -0.0231 | -0.130 | -0.103 | -0.0420 | -0.0932 |
| | (0.0743) | (0.0729) | (0.0792) | (0.0651) | (0.0746) | (0.0689) |
| Drinking alcohol | -0.112*** | -0.165*** | -0.119*** | -0.138*** | -0.0550** | -0.199*** |
| | (0.0243) | (0.0333) | (0.0310) | (0.0243) | (0.0257) | (0.0292) |
| Sport practice | 0.0316 | 0.0156 | 0.0740** | -0.0291 | 0.0865*** | -0.0414 |
| | (0.0313) | (0.0358) | (0.0376) | (0.0298) | (0.0320) | (0.0343) |
| Self-Assessed-Health (discrete) | 0.281 | -0.279 | -0.172 | 0.523*** | 0.246 | 0.173 |
| | (0.195) | (0.407) | (0.274) | (0.203) | (0.248) | (0.252) |
| Self-Assessed-Health (good) | 0.0322 | -0.542 | -0.251 | 0.191 | -0.00170 | -0.0535 |
| | (0.211) | (0.409) | (0.291) | (0.218) | (0.262) | (0.276) |
| Self-Assessed-Health (very good) | 1.099** | -0.232 | -1.005*** | 1.148*** | 0.250 | 0.435 |
| | (0.445) | (0.454) | (0.374) | (0.294) | (0.397) | (0.350) |
| Self-Assessed-Health (excellent) | 0.426* | -0.324 | -0.195 | 0.613*** | 0.231 | 0.268 |
| | (0.237) | (0.415) | (0.305) | (0.229) | (0.275) | (0.291) |
| Regional fixed effects | Yes | Yes | Yes | Yes | Yes | Yes |
| /cut1 | 0.0316 | 0.0156 | 0.0740** | -0.0291 | 0.0865*** | -0.0414 |
| | (0.0313) | (0.0358) | (0.0376) | (0.0298) | (0.0320) | (0.0343) |
| /cut2 | 0.281 | -0.279 | -0.172 | 0.523*** | 0.246 | 0.173 |
|          | (0.195)  | (0.407)  | (0.274)  | (0.203)  | (0.248)  | (0.252)  |
| /cut3    | 0.0322   | -0.542   | -0.251   | 0.191    | -0.00170 | -0.0535  |
|          | (0.211)  | (0.409)  | (0.291)  | (0.218)  | (0.262)  | (0.276)  |
| /cut4    | 1.099**  | -0.232   | -1.005***| 1.148*** | 0.250    | 0.435    |
|          | (0.445)  | (0.454)  | (0.374)  | (0.294)  | (0.397)  | (0.350)  |
| Log likelihood |    |          |          |          |          |          |
| Adj. R Squared |    |          |          |          |          |          |
| Observations | 2,567 | 2,792 | 2,157 | 3,202 | 2,756 | 2,603 |

Omitted benchmark: lowest income level, student, single, no children, primary education, self-assessed-health: bad. Robust standard errors in parentheses. *** p<0.01, ** p<0.05, * p<0.1.